# Gate-defined Quantum Point Contact in an InSb Two-dimensional Electron Gas

Zijin Lei[*], Christian A. Lehner, Erik Cheah, Christopher Mittag, Matija Karalic,

Werner Wegscheider, Klaus Ensslin, and Thomas Ihn

*Solid State Physics Laboratory, ETH Zurich, CH-8093 Zurich, Switzerland*

[*]*Email of the corresponding author: zilei@phys.ethz.ch*

**Abstract**

We investigate an electrostatically defined quantum point contact in a high-mobility InSb two-dimensional electron gas. Well-defined conductance plateaus are observed, and the subband structure of the quantum point contact is extracted from finite-bias measurements. The Zeeman splitting is measured in both in-plane and out-of-plane magnetic fields. We find an in-plane *g* factor $|g_\parallel^*| \approx 40$. The out-of-plane *g* factor is measured to be $|g_\perp^*| \approx 50$, which is close to the *g* factor in the bulk.

___________________________

Indium antimonide (InSb) is a III-V binary compound known for its low effective mass, giant effective *g* factor in the bulk, and its large spin-orbit interactions (SOIs) [1-5]. These unique properties are interesting in view of applications such as high-frequency electronics [6], optoelectronics [7], and spintronics [8]. Recently, InSb, as well as InAs, has received more and more attention as a candidate to realize Majorana zero modes at the boundary to topological superconductors [9]. The topological superconducting phase can be achieved by combining superconductivity induced by proximity effects, control of the Zeeman field, a strong Rashba SOI, and the phase of Josephson junctions or SQUIDs if the system is two-or three dimensional [9-13]. An in-depth understanding of InSb nanostructures is necessary for scaling up and integrating potential InSb-based topological quantum devices. Investigating the Zeeman effect and SOIs at the nanoscale will help us to understand the topological nontrivial phase achieved in Majorana nanodevices. A quantum point contact (QPC) is one of the basic nanostructures, where a ballistic charged-carriers system is confined into a one-dimensional channel. In such a structure, the conductance of the channel is quantized in integer multiples of the conductance quantum $e^2/h$ [14-16]. Therefore, the conductance $e^2/h$ is called conductance quantum as a natural unit. Conductance quantization has been found in InSb nanostructures based on as-grown nanowires [17-20], nano sails [21], and other free-standing nanomaterials [22, 23]. However, reports on InSb QPCs defined in InSb quantum wells (QWs) are still rare. Conductance quantization in InSb QWs is difficult to achieve [24-27], even though high-mobility InSb QWs can nowadays be grown with molecular beam epitaxy (MBE) [28, 29] and various transport experiments have been performed [30-34]. While carrier mobility is high (several 100,000 cm$^2$/Vs) and therefore the elastic mean free path easily exceeds the dimensions of the quantum point contacts, time-dependent shifts of the device characteristics lead to serious hysteresis effects when sweeping the gate voltages. This is the main obstacle for high-quality InSb-QW-based QPCs and other nanostructures such as quantum dots. Due to this time-dependent effect, it is challenging to deplete the two-dimensional electron gas (2DEG) in the QW in a stable way for a sufficiently long period of time, during which a transport experiment can be performed. The reason for the time-dependent effect may be related to the Si-doping in the InAlSb barriers, from which the electrons in the QWs originate [35]. In previous works, a chemical etching method was adopted to define a one-dimensional channel, and a metal top gate or a pair of side gates were used to tune the density in the channel locally to achieve a stable pinch-off [26, 27]. As a comparison provided in the Supplementary Material of Ref. 26, a gate-defined QPC in an InSb QW was characterized to be inferior in quality. The etching method seriously limits the device quality due to the induced scattering centers at the edges of the fabricated structures. Furthermore, the etching step makes the design and processing of integrated nanodevices much more challenging as compared to gate-defined devices. Recently, Kulesh *et. al* [35] reported a purely gate-defined stable quantum dot on InSb QWs. To solve the problem of hysteresis, an undoped InSb QW was used and an extra global top gate was added to induce electrons into the QW electrostatically. This undoped InSb may have higher mobility due to the absence of the remote ionized scattering centers. Nevertheless, the potential of this kind of QWs still needs to be investigated in further experiments.

In this work, we study a QPC defined electrostatically in an InSb 2DEG populated by remote doping. We provide a detailed characterization of its energy levels, magnetoelectric subband structure, and effective *g* factor with the magnetic field applied in different directions relative to the plane of the 2DEG. With a special procedure for sweeping the gate voltage, the QPC can be dynamically stabilized, enabling successive measurements with nearly identical electronic conditions.

Figure 1 (a) shows the layer sequence of the sample. The InSb QW containing the two-dimensional electron gas on which the QPC is defined is grown on a (100) GaAs substrate by MBE. An interfacial misfit transition to the GaSb buffer and an interlayer InAlSb buffer is employed to overcome the lattice mismatch between GaAs and InSb. The total thickness of the buffer layers amounts to roughly 3 $\mu$m. The 21 nm-thick InSb QW is surrounded by $In_{0.9}Al_{0.1}Sb$ confinement barriers, the n-type carriers are introduced to the QW by two Si δ-doping layers incorporated in the barriers, 40 nm below and above the QW, respectively. On the top of the QW, an $In_{0.9}Al_{0.1}Sb$ layer with a thickness of 100 nm is grown. More details about the MBE growth can be found in Ref. 29.

Figure 1 (b) shows a schematic diagram of the QPC gate structure and the measurement setup. The QPC here is defined on a standard Hall bar structure (light grey in the figure). The fabrication process of the Hall bar is similar to our previous work [34]. First, a Hall bar structure with the size of 400× 200 μm² is defined by wet chemical etching with an etching depth of more than 160 nm, which is deeper than the Si δ-doping layer on the substrate side. Second, layers of Ge/Ni/Au are evaporated onto the contact areas after Ar milling. In the next step, the sample is coated with a 40 nm thick aluminum oxide (ALO) dielectric layer using atomic layer deposition (ALD) at a temperature of 150 °C. A high-temperature annealing step is unnecessary because the metal diffuses into the material during the ALD process which heats the sample. Finally, pairs of split gates are deposited onto the ALO layer. To avoid a potential interruption of the gates of the gate-metallization at the mesa edge, we first fabricate the inner thin (5/25 nm) Ti/Au nanometer-sized gates (dark grey in the figure) on top of the mesa with standard electron beam lithography followed by electron beam evaporation. Then, we use optical lithography and electron beam evaporation to define the thick (10/120 nm) Ti/Au micrometer-sized gate patches to connect the fine gates across the mesa edge to contact pads outside the mesa structure. The QPC measured in this work has a split-gate separation of 200 nm. Before further studying the QPC, we characterize the 2DEG with standard magneto-transport experiments at 1.3 K using the Hall bar geometry of the sample with all gates grounded. The mobility of the 2DEG is $\mu = 1.04 \times 10^5$ cm$^2$/(Vs) and the electron density $n = 1.3 \times 10^{15}$ m$^{-2}$. Based on these numbers, we estimate the mean free path to be $l_e = 620$ nm, which is larger than the split gate separation and the lithographic channel length of 500 nm.

The transport measurement of the QPC uses standard low frequency (5 Hz) lock-in techniques together with a DC measurement in a He$^4$ cryostat with a base temperature of 1.3 K. The same DC bias $V_{sg}$ is applied to both parts of the split-gate to form the QPC in the 2DEG. We apply a fixed AC bias of $V_{AC} = 350$ μV and a variable DC bias $V_{bias}$ between contacts 1 and 4 and measure the AC and DC components of the two-terminal current $I_{AC}$ and $I_{DC}$. In addition, we measure the AC and DC parts of the diagonal voltage-drop $V_{36,AC}$ and $V_{36,DC}$ between contacts 3 and 6. A rotatable magnetic field $B$ can be applied in our experiment, where the angle between the direction of the magnetic field and the sample normal can be precisely calibrated with the Hall measurement. As shown in Fig. 1 (b), we denote the magnetic field applied perpendicular to the sample surface as $B_\perp$, and the magnetic field aligned parallel to the sample surface but perpendicular to the current as $B_\parallel$.

The QPC is created by applying a negative $V_{sg}$. A special sweeping protocol is adopted for the gate voltages to achieve reproducible measurement results. The 1D channel can be pinched off completely by a sufficiently negative split gate voltage. However, due to a time-dependent shift of the gate voltage characteristic in our Si-doped InSb/InAlSb heterostructures, the pinch-off lasts less than a minute in our measurement. This short time scale does not allow us to perform successive conductance measurements for different source-drain voltages or magnetic fields. We, therefore, loop the gate voltage $V_{sg}$ continuously between two carefully chosen bounds and thereby achieve reproducible gate characteristics that are stable in time. For all the measurements shown below, the protocol for sweeping $V_{sg}$ is as follows: We measure the first $V_{sg}$-dependent conductance trace from zero voltage to channel pinch-off. Without any delay, we then tune $V_{sg}$ back to zero and keep it there for at least 5 minutes. The next traces are obtained by repeating this procedure. We find that different sweep rates and different resting times at $V_{sg} = 0$ can change the $V_{sg}$ required to pinch off the channel. Therefore, all the measurements shown below are performed with the same measurement protocol within one cool-down with a total measurement time of three weeks.

Figure 2 (a) presents the differential conductance $G_{\text{diff}} = I_{\text{AC}}/V_{36,\text{AC}}$ as a function of $V_{\text{sg}}$ when $V_{\text{bias}} = 0$. A constant series resistance $R_s$ = 1.2 kΩ is subtracted. In the later experiment, $R_s$ is always kept at 1.2 kΩ, assuming that the variation of the magnetoresistance is negligible in the range of $B$ where the measurement is performed. With decreasing $V_{\text{sg}}$, the channel gradually gets pinched off and conductance steps that correspond to 6, 4, and 2 conductance quanta can be observed. The absolute heights of these steps are always lower than the correct values, possibly because of backscattering, and the curve is more oscillatory than for a standard QPC in GaAs heterostructures.

In the next step, we perform finite bias spectroscopy by applying a nonzero $V_{\text{bias}}$ to measure the mode spacing of the QPC. With the measurement protocol introduced above, we measure the $V_{\text{sg}}$ dependence of the conductance, and step $V_{\text{bias}}$ after each such sweep while $V_{\text{sg}}$ is zero. This is different from the commonly used way in which $V_{\text{bias}}$ is swept and $V_{\text{sg}}$ is stepped. Here, we subtract the voltage drop across $R_s$ from $V_{36,\text{DC}}$ to obtain $V_{\text{DC}}$, i.e., $V_{\text{DC}} = V_{36,\text{DC}} - I_{\text{DC}} \times R_s$. Figure 2 (b) depicts the differential transconductance $dG_{\text{diff}}/dV_{\text{sg}}$ of the QPC as a function of both $V_{\text{sg}}$ and $V_{\text{DC}}$. The dark regions in the colormap represent conductance plateaus, the light regions mark the transition between them. The extent in $V_{\text{DC}}$ of the diamond-like plateau regions measures the energy separation of the QPC modes. We read the mode spacing using the auxiliary green dotted lines in the figure and find values $\Delta E_{1,2} \approx$ 3.8 meV and $\Delta E_{2,3} \approx$ 3.5 meV. In a harmonic potential approximation, this mode spacing is related to the frequency $\omega_0$ via $\Delta E_{i,i+1} = \hbar\omega_0$. The real-space extents of the modes $L_n$ with $n$ = 1 and $n$ = 2 are then calculated according to $\frac{m^*}{2}\omega_0^2 L_n^2 = \hbar\omega_0(n - \frac{1}{2})$. Here, we use the electron effective mass $m^*$ = 0.017 $m_e$, where $m_e$ is the free electron mass. This value has been obtained through the temperature dependence of Shubnikov-de Haas oscillations in our previous work on InSb QWs with the same thickness [34]. We find $L_1 \approx$ 34 nm and $L_2 \approx$ 64 nm, as expected smaller than the separation of the split-gates.

In the following, we investigate the effect of $B_\perp$ on transport through the QPC. In Fig. 3 (a), the dependence of $G_{\text{diff}}$ on $V_{\text{sg}}$ for different $B_\perp$ is presented. With increasing $B_\perp$, the absolute heights of the conductance plateaus gradually move to the expected quantized values, and the step-like features become more pronounced. This is because the applied $B_\perp$ reduces backscattering through the channel. Due to the Zeeman effect, the two spin-degenerate states in each mode start to separate and the plateaus at odd multiples of the quantum conductance become observable. The transconductance $dG_{\text{diff}}/dV_{\text{sg}}$ is presented as a function of $B_\perp$ and $V_{\text{sg}}$ in Fig. 3 (b). For increasing $B_\perp$ the bright lines, which indicate the transitions between conductance plateaus, curve towards higher gate voltages, an effect known as the magnetic depopulation of QPC modes. They tend to approach a linear slope as they gradually merge into the Landau levels which form at high magnetic fields. The magnitude of the Zeeman energy can be mapped by finite bias measurement with a nonzero $B_\perp$ applied. Figure 3 (c) shows the transconductance $dG_{\text{diff}}/dV_{\text{sg}}$ vs. $V_{\text{DC}}$ and $V_{\text{sg}}$ at $B_\perp = 1.15$ T. The extent in $V_{\text{DC}}$ direction of the first and the third diamonds correspond to the Zeeman energy $\Delta E_{1\uparrow\downarrow}$ and $\Delta E_{2\uparrow\downarrow}$ between the spin-polarized states where $n = 1$ and $n = 2$, respectively. As presented in Fig. 3 (d), by repeating the finite bias measurement with different $B_\perp$, we can estimate the out-of-plane effective $g$ factor $g^*_\perp$ to have the value $|g^*_\perp| \approx 50$ with an uncertainty of about 10%, by linearly fitting $\Delta E_{1\uparrow\downarrow}$ and $\Delta E_{2\uparrow\downarrow}$ vs. $B_\perp$.

In addition, we rotate the sample to have the magnetic field $B_\parallel$ applied in-plane, but perpendicular to the current. Figure 4 shows the transconductance $dG_{\text{diff}}/dV_{\text{sg}}$ as a function of $V_{\text{sg}}$ and $B_\parallel$. Each spin-degenerate state which contributes a conductance of $2e^2/h$ at zero field, gradually splits into two spin-polarized states with increasing $B_\parallel$. As shown in Fig. 5 (a), which is a cut of Fig. (4) at $B_\parallel = 0.75$ T (white dashed line), conductance plateaus with both even and odd multiples of the conductance $e^2/h$ are visible. This is verified in the transconductance curve and the finite bias measurement presented in Fig. 5 (b), where the number of the transconductance minima in the given range of $V_{\text{sg}}$ doubles as compared to zero field. Increasing $B_\parallel$ in the range 1.3 T $< B_\parallel <$ 1.9 T in Fig. 4, we can tune the energy separation of two spin-polarized states in one subband to be about equal to the zero-field subband separation. For instance, as presented in Fig. 5(c), which is a cut along at $B_\parallel = 1.32$ T in Fig. 4 (green dashed line), the conductance plateaus now occur at odd multiples of $e^2/h$ only. This is also seen in the transconductance trace and the finite bias measurement shown in Fig. 5(d), where the mode spacing is close to its zero-field value. Increasing $B_\parallel$ further, the Zeeman energy increases continuously. This leads to a pattern of closing and reopening gaps between neighboring spin-polarized states in Fig. 4 as indicated by the numbers.

The in-plane effective $g$ factor $|g_\parallel^*|$ can be estimated by combining the measurements performed with $B_\parallel$ applied. There are two important features to be noticed. First, the lever arm of the split gate, relating a change of $V_{sg}$ to a shift in mode energy, is nearly constant when $V_{sg} < -1$ V. This is found from the similarity of the slopes of the green dotted lines in Figs. 5 (b) and (d). Second, in the same figures, the green dotted lines with the same direction are nearly parallel. These two observations indicate that the subband spacing $\Delta E_{1,2}$ is independent of $B_\parallel$. We examine the height of the transconductance peak around $V_{sg}$ = -1.05 V by cutting Fig. 4 horizontally at different values of $B_\parallel$. We find that the Zeeman energy most closely equals the energy separation between the modes with $n = 1$ and $n = 2$ when $B_\parallel$ = 1.65 T, i.e.,

$$|g_\parallel^*| \mu_B B_\parallel|_{B_\parallel = 1.65 \text{ T}} = \Delta E_{1,2},$$

where $\mu_B$ is the Bohr magneton. With the value $\Delta E_{1,2} \approx 3.8$ meV extracted from Fig. 2 (b), we estimate the in-plane effective $g$ factor to be $|g_\parallel^*| \approx 40$ with an uncertainty of about 10%, which is lower than $|g_\perp^*|$ found before. Alternatively, $|g_\parallel^*|$ can be extracted from the finite bias measurement presented in Fig. 5 (b). Similar to the situation where only $B_\perp$ is applied, the extent of the diamonds with conductance $e^2/h$ and $3e^2/h$ in $V_{DC}$ direction corresponds to the Zeeman energy when $B_\parallel$ = 0.75 T, where the Zeeman energy is observable but still smaller than $\Delta E_{1,2}$. Thus, we can write $|g_\parallel^*| \mu_B B_\parallel = \Delta E_{1\uparrow\downarrow}$ and $|g_\parallel^*| \mu_B B_\parallel = \Delta E_{2\uparrow\downarrow}$. With the measurement showing $\Delta E_{1\uparrow\downarrow} \approx \Delta E_{2\uparrow\downarrow} \approx 1.8$ meV, we estimate $|g_\parallel^*| \approx 41$. Due to limited data quality, this estimation may have a larger error than the estimate obtained from the zero bias measurement shown in Fig. 4. Furthermore, there may be larger errors in the estimation of $|g_\parallel^*|$ as compared to that of $|g_\perp^*|$ because the parallel magnetic field does not reduce backscattering.

We now discuss our measurement results in the light of previous publications. Our device delivered a better data quality than the gate-defined device presented in Ref. 26. Comparing our results with devices defined by etching, we find a slightly lower anisotropy of the effective *g* even though experimental uncertainties are significant. In Ref. 26 and Ref. 27 defined by chemical-etching, the in-plane effective *g* factor was found to be about half of the out-of-plane effective *g* factor. In Ref. 26, the in-plane *g* factor is ~ 26 but the out-of-plane *g* factor is ~ 52. Similarly, in Ref. 27, the in-plane *g* factor is ~ 40 while the out-of-plane *g* factor is ~ 60. The lower *g* factor anisotropy in our device may be related to a weaker SOI in our system. Because the Si-doping is incorporated almost symmetrically around the QW in our device, we expect the wavefunction of electrons to sit more symmetrically in the QW than in previous works. Therefore, the Rashba contribution to the SOI is likely to be smaller in our system. Because of the huge effective *g* factor in the bulk, the Zeeman energy can exceed the contribution of the Dresselhaus SOI at a relatively small magnetic field. In agreement with that, within our measurement precision, a smaller *g* factor anisotropy is observed. As a result of the insignificant SOI and the data quality limitation, we did not find signatures of avoided-crossing effects of the levels either. The tendency that the out-of-plane *g* factor is larger than the in-plane *g* factor is in accordance with the calculation in Ref. 36, but both *g* factors should not exceed the *g* factor in the bulk if only the **k·p** model is adopted. In our device, the value $|g^*_\perp| \approx 50$ is close to the *g* factor in the bulk but much larger than the value obtained from the coincidence measurement. A detailed introduction of the coincidence measurement and the **k·p** theory calculation can be found in Ref. 30, 31, and 34. Especially, in Ref. 34, the *g* factor of an InSb QW with the same thickness shows a value of $|g^*| = 35$ and it meets the **k·p** theory result, when an increased bandgap due to the QW confinement is considered. We were able to reproduce this coincidence measurement result with the Hall bar of the QPC-device presented in this paper finding the *g* factor to be between 35 and 40. Furthermore, we find that there is no obvious *g* factor enhancement due to electron-electron interactions. Beyond the comparison to QWs, the effective *g* factor that we obtained from the QPC measurements may be compared with *g* factor measurements in QPCs based on InSb nanowires. Although in most of the publications [18, 19] the effective *g* factor is found to be around 40 because of the confinement, it is possible that the effective *g* factor is enhanced in these devices due to electron-electron interactions in the constriction, where the carrier density is low [17]. However, since there is no observable signature related to strong electron-electron interactions, such as the 0.7 $e^2/h$ anomaly, this interpretation still needs more support by investigating QPCs with higher quality.

In conclusion, we presented a completely gate-defined QPC device based on a two-dimensional electron gas in an InSb QW. With a special gate operation protocol, the device maintains dynamical electrostatic stability. The energy separations between the QPC modes are determined using finite bias measurements. Spin-resolved transport through the nanostructure is observed in both in-plane and out-of-plane magnetic field. The value of the effective *g* factor is ~ 50 for the magnetic field applied normal to the plane and ~ 40 with the field applied in-plane. The out-of-plane *g* factor is larger than the value obtained from the 2DEG using the coincidence method in the same device, and larger than the value estimated from ***k·p*** theory. The unusual but necessary measurement protocol limits the further development of InSb nanodevices because of the long measurement time and the challenge of integrating more gates. Therefore, the problem of time-dependent gate characteristics needs to be solved before further progress can be made. We expect that more optimization of the heterostructure growth will provide enhanced device stability, which may pave the way for more elaborate nanostructures based on InSb QWs.

We thank Dr. F. K. de Vries and Mr. L. Ginzburg for fruitful discussions. We thank Mr. P. Märki and Mr. T. Bähler for their technical support. This work was supported by the Swiss National Science Foundation through the National Center of Competence in Research (NCCR) Quantum Science and Technology.

**Figure 1.** (a) Layer structure of the QW heterostructure. (b) A schematic representation of the QPC. The QPC is defined on a standard Hall bar geometry (light gray). The gates of the QPC (dark gray) are evaporated on the ALOx dielectric layer. The gates separation is 200 nm. The in-plane and out-of-plane magnetic field $B_\parallel$ and $B_\perp$ are defined as illustrated. The current and voltage are measured in the configuration presented here as well.

**Figure 2.** (a) Differential conductance $G_{\text{diff}}$ as a function of $V_{\text{sg}}$ at a temperature of $T = 1.3$ K when no magnetic field is applied. Steps of the conductance are visible. A series resistance $R_s = 1.2$ k$\Omega$ has been subtracted. $G_{\text{diff}} = 2e^2/h$, $4e^2/h$, and $6e^2/h$ are labeled with dashed lines. (b) Finite bias spectroscopy showing the transconductance $dG_{\text{diff}}/dV_{\text{sg}}$ as a function of $V_{\text{sg}}$ and $V_{\text{DC}}$. A correction on the voltage drop through the QPC is made with $R_s = 1.2$ k$\Omega$. The green dashed lines are added as guidance.

**Figure 3.** (a) Differential conductance as a function of both, $B_\perp$ and $V_{\text{sg}}$. The plateaus move to the correct value with the increase of $B_\perp$, and spin-split conductance plateaus are observable. A series resistance $R_s = 1.2$ k$\Omega$ has been subtracted. $G_{\text{diff}} = 2e^2/h$ and $4e^2/h$ are labeled with dashed lines. (b) Transconductance $dG_{\text{diff}}/dV_{\text{sg}}$ as a function of $B_\perp$ and $V_{\text{sg}}$. Both magnetic depopulation and spin splitting are visible. In different dark regions, the corresponding values of the conductance are labeled in the unit of $e^2/h$. (c) Finite bias spectroscopy showing the transconductance $dG_{\text{diff}}/dV_{\text{sg}}$ as a function of $V_{\text{sg}}$ and $V_{\text{DC}}$ when $B_\perp = 1.15$ T. The green dashed lines are added as guidance. (d) With a series reproduction of measurement in (c), the $B_\perp$ dependence of $\Delta E_{1\uparrow\downarrow}$ and $\Delta E_{2\uparrow\downarrow}$ is obtained. The error bars are determined by observing the height of the transconductance peaks in the cuts in the $V_{\text{sg}}$ direction. The linear fit shows the out-of-plane g factor with a value of $|g^*_\perp| \approx 50$.

**Figure 4.** Transconductance $dG_{\text{diff}}/dV_{\text{sg}}$ as a function of $B_\parallel$ and $V_{\text{sg}}$. The spin-degenerated conductance plateaus split according to the Zeeman effect and are labeled with the associated conductance in the unit of $e^2/h$.

**Figure 5.** (a) and (c) The $V_{\text{sg}}$ dependence of the differential conductance $G_{\text{diff}}$ and transconductance $dG_{\text{diff}}/dV_{\text{sg}}$ when in-plane magnetic fields $B_\parallel = 0.75$ T and $B_\parallel = 1.32$ T are applied respectively. The conductance plateaus are labeled in unit of $e^2/h$. (a) and (c) are the cuts of Fig. 4 along the white and green dashed line respectively. (b) and (d) are finite bias spectroscopy showing the transconductance $dG_{\text{diff}}/dV_{\text{sg}}$ as a function of $V_{\text{sg}}$ and $V_{\text{DC}}$ with $B_\parallel = 0.75$ T and $B_\parallel = 1.32$ T applied, respectively. Green dashed lines are guidance.

**Figure 1** (*this is a 1-column figure*)

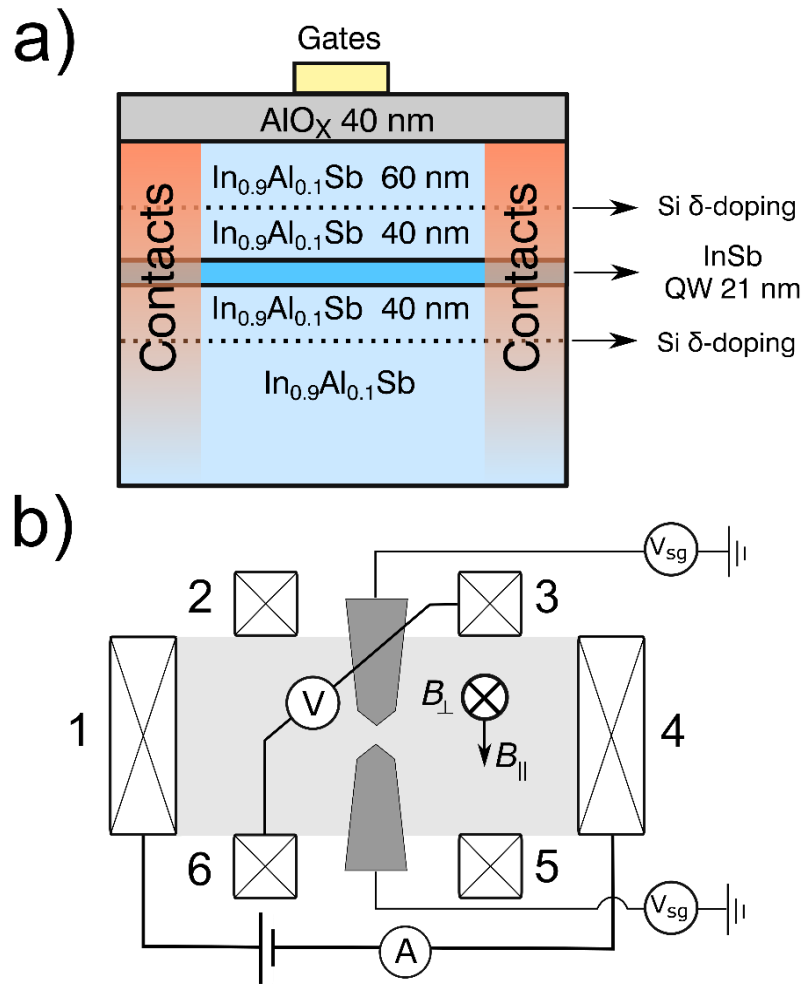

**Figure 2** (*this is a 1-column figure*)

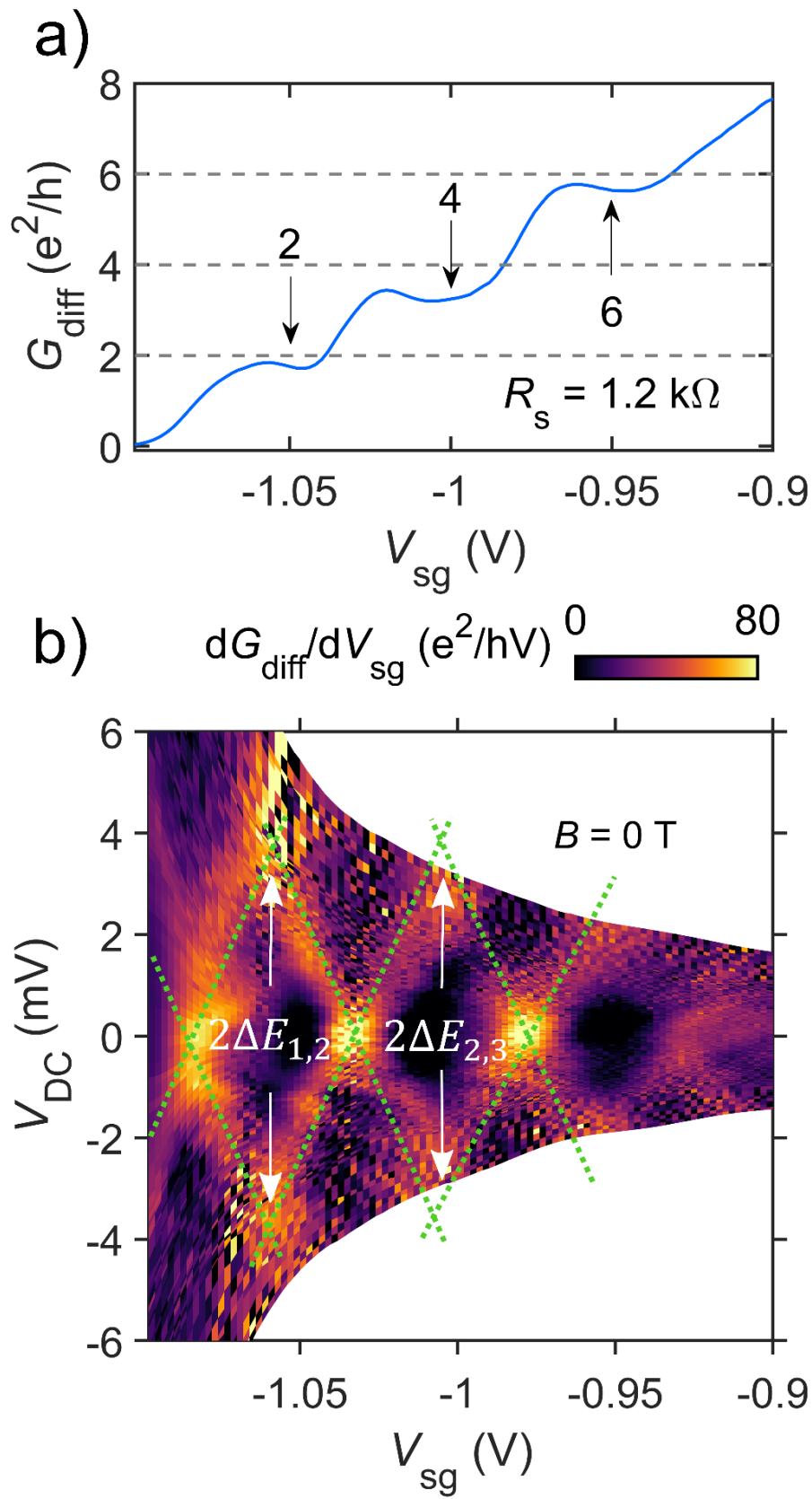

**Figure 3** (*this is a 2-column figure*)

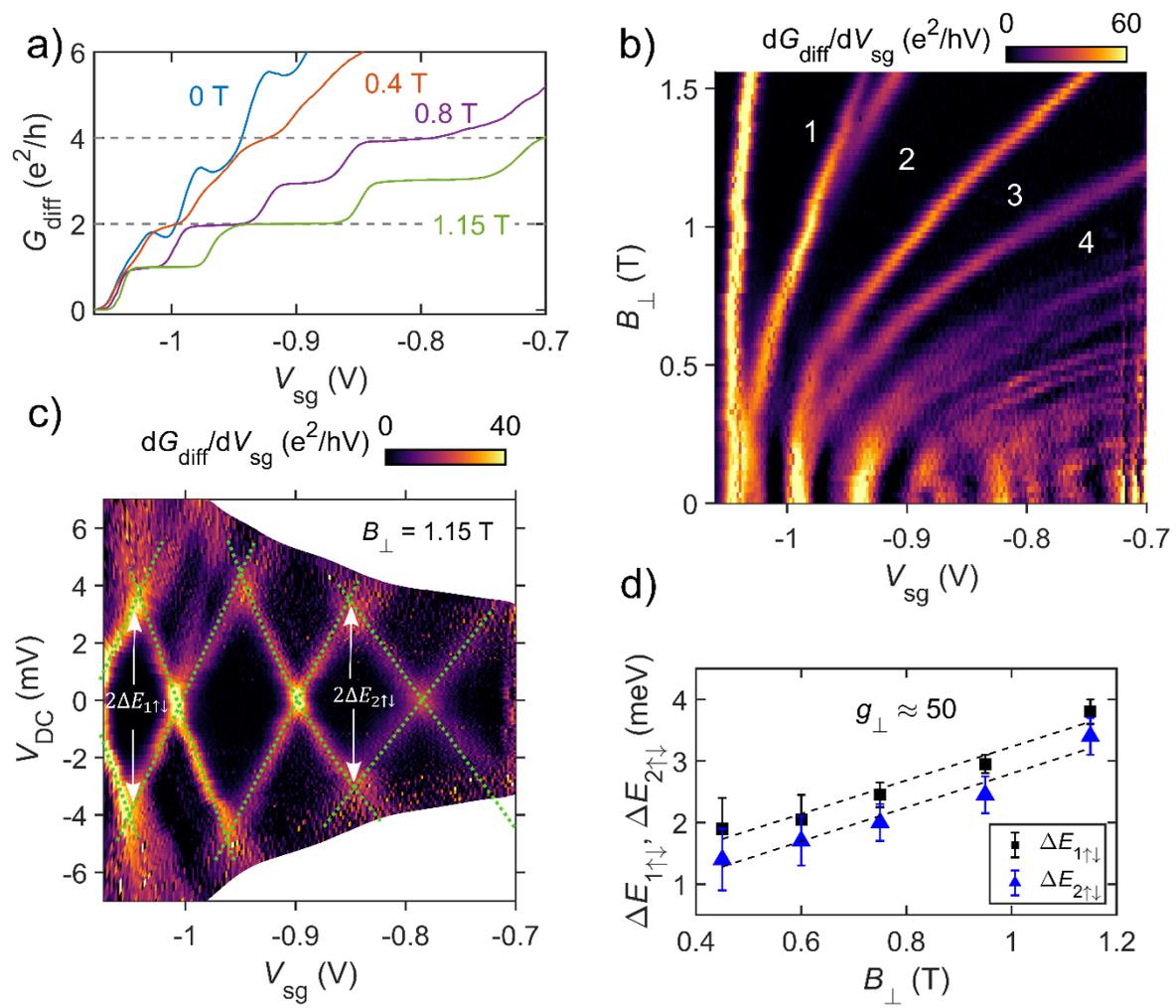

**Figure 4** (*this is a 1-column figure*)

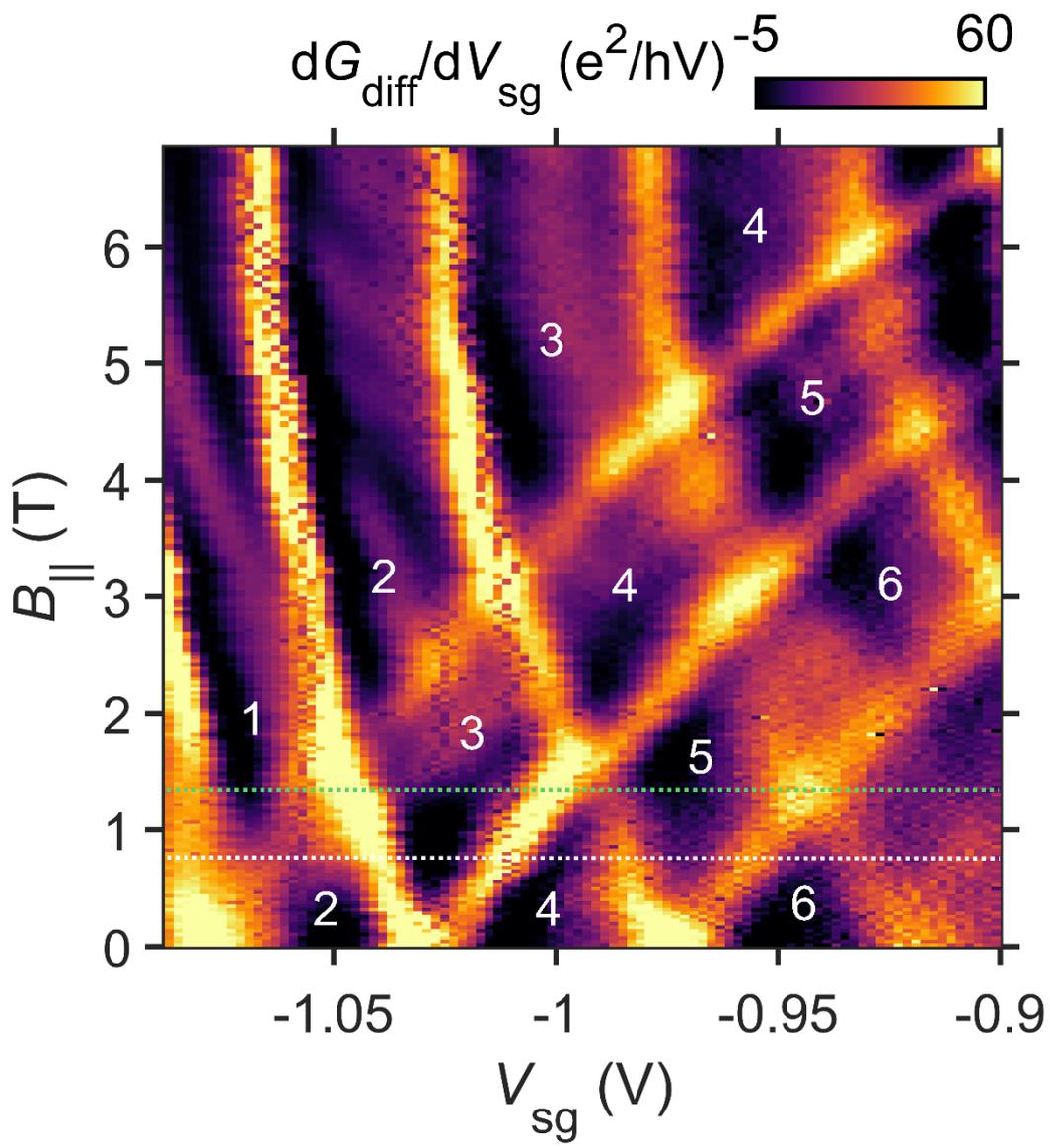

**Figure 5** (*this is a 2-column figure*)

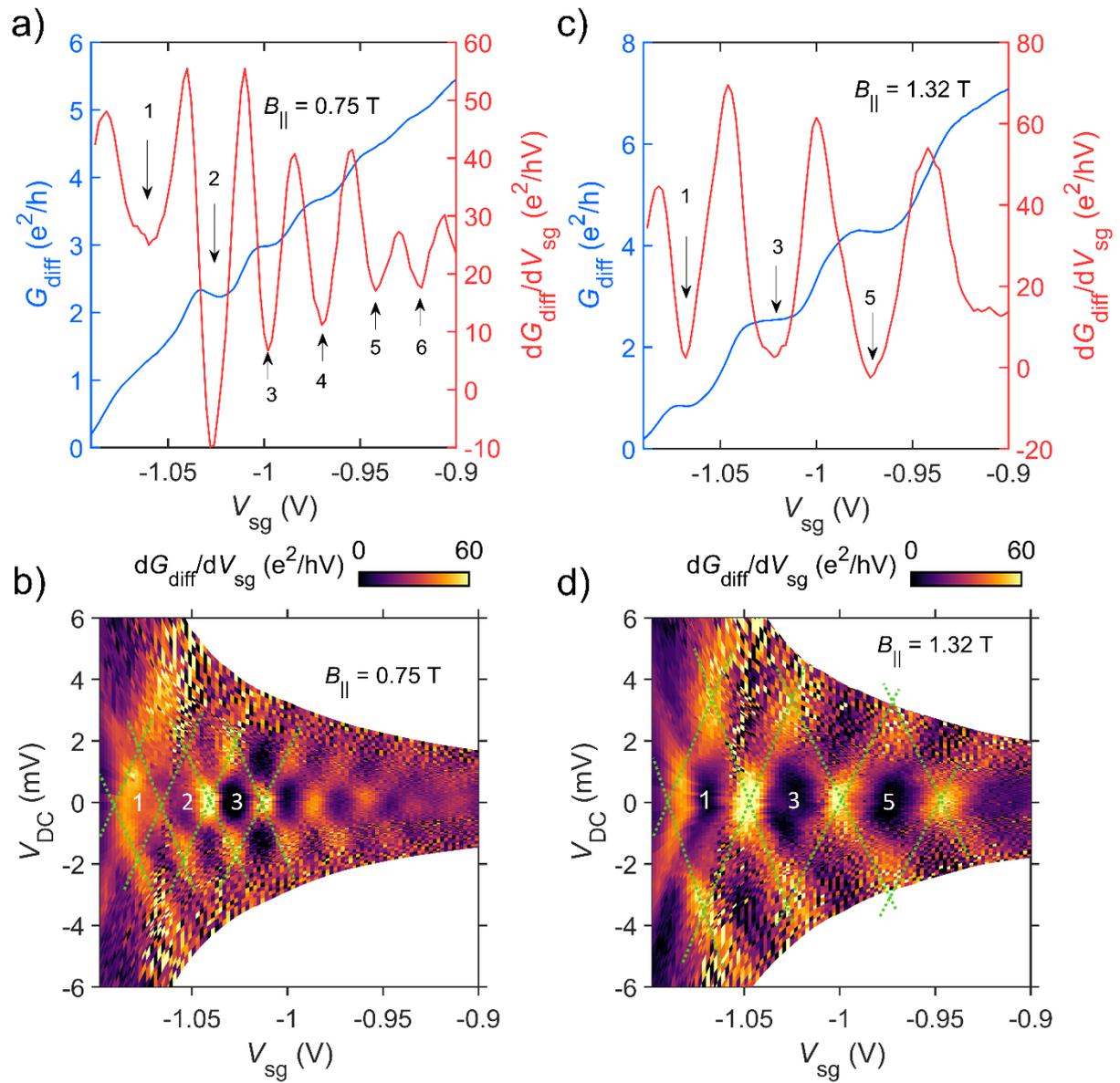